\documentclass[conference]{IEEEtran}
\IEEEoverridecommandlockouts
\usepackage{cite, soul}
\usepackage{amsmath,amssymb,amsfonts}
\usepackage{algorithmic}
\usepackage{graphicx}
\usepackage{textcomp}
\usepackage{tikz}
\usepackage{enumitem}
\usepackage{forest}
\usepackage{xcolor}
\usepackage{caption}
\usepackage{subcaption}
\usepackage{balance}

\newcommand{\linebreakand}{
  \end{@IEEEauthorhalign}
  \hfill\mbox{}\par
  \mbox{}\hfill\begin{@IEEEauthorhalign}
}

\begin{document}

\title{Overcoming DNSSEC Islands of Security: \\ A TLS and IP-Based Certificate Solution}

% \author{\IEEEauthorblockN{1\textsuperscript{st} Rishith Aduma}
% \IEEEauthorblockA{\textit{CSE. IIT Dharwad)} \\
% \and
% \and

\author{\IEEEauthorblockN{Aduma Rishith\textsuperscript{*}, Aditya Kulkarni\textsuperscript{*}, Tamal Das\textsuperscript{*}, and Vivek Balachandran\textsuperscript{\#}}
\IEEEauthorblockA{\textsuperscript{*}\emph{Indian Institute of Technology, Dharwad, India} \\
\textsuperscript{\#}\emph{Singapore Institute of Technology, Singapore} \\
\textsuperscript{*}\{210010002, aditya.kulkarni, tamal\}@iitdh.ac.in, \textsuperscript{\#}vivek.b@singaporetech.edu.sg}}

\maketitle

\begin{abstract}
The Domain Name System (DNS) serves as the backbone of the Internet, primarily translating domain names to IP addresses. Over time, various enhancements have been introduced to strengthen the integrity of DNS. Among these, DNSSEC stands out as a leading cryptographic solution. It protects against attacks (such as DNS spoofing) by establishing a chain of trust throughout the DNS nameserver hierarchy. However, DNSSEC's effectiveness is compromised when there is a break in this chain, resulting in ``Islands of Security'', where domains can authenticate locally but not across hierarchical levels, leading to a loss of trust and validation between them.
Leading approaches to addressing these issues were centralized, with a single authority maintaining some kind of bulletin board. This approach requires significantly more infrastructure and places excessive trust in the entity responsible for managing it properly.

In this paper, we propose a decentralized approach to addressing gaps in DNSSEC's chain of trust, commonly referred to as ``Islands of Security". We leverage TLS and IP-based certificates to enable end-to-end authentication between hierarchical levels, eliminating the need for uniform DNSSEC deployment across every level of the DNS hierarchy. This approach enhances the overall integrity of DNSSEC, while reducing dependence on registrars for maintaining signature records to verify the child nameserver's authenticity. By offering a more flexible and efficient solution, our method strengthens DNS security and streamlines deployment across diverse environments.
\end{abstract}


% \section{Introduction}
% This document is a model and instructions for \LaTeX.
% Please observe the conference page limits. 

\section{Introduction}

The \textit{Domain Name System (DNS)} is a vital component of the Internet, translating human-readable domain names (e.g., \texttt{www.example.com}) into machine-readable IP addresses (e.g., \texttt{192.0.2.1})~\cite{rfc1035}. However, since its inception, DNS has lacked mechanisms for verifying the authenticity and integrity of responses. This vulnerability has led to various forms of attack, including \textit{DNS cache poisoning}~\cite{rfc3833}, where attackers can inject falsified responses into the cache of recursive resolvers, redirecting users to malicious websites. In light of these issues, the need for stronger security measures became clear.

One proactive solution -- named \textit{Domain Name System Security Extensions (DNSSEC)} -- was introduced in the early 2000s~\cite{rfc4033}. DNSSEC provides cryptographic validation of DNS responses by enabling digital signatures for DNS records, ensuring their integrity. By establishing a \textit{chain of trust} through the DNS hierarchy, DNSSEC protects against tampering and ensures the integrity of DNS data. Despite its effectiveness, DNSSEC adoption has been slow, leaving many infrastructures unprotected and vulnerable to classical attacks like cache poisoning.

In the absence of widespread DNSSEC deployment, in 2008, security researcher Dan Kaminsky uncovered a critical weakness in DNS, leading to a more effective form of cache poisoning known as the \textit{Kaminsky attack}~\cite{alexiou2010formal}. By exploiting the predictability of DNS transaction IDs and source ports (explored in Section-~\ref{DNS_section}), attackers could rapidly inject fraudulent DNS responses. To counter this, \textit{port randomization} was introduced, making it significantly harder for attackers to guess the correct combination of transaction ID and source port~\cite{rfc6056}. Additional mitigations like \textit{bailiwick checking}~\cite{rfc8499} and \texttt{0x20 encoding}~\cite{dagon2008increased} further enhanced security by restricting the domains a nameserver could provide responses for and adding randomness by varying the case of letters in the DNS queries, respectively. 

Despite these defenses, the integrity of DNS responses remained a concern. Over the next decade, attackers continued to exploit DNS vulnerabilities such as \textit{DNS hijacking} and \textit{DNS injection}, which became common techniques where attackers manipulated DNS responses. While these attacks didn’t directly poison caches, they targeted the integrity of DNS responses by altering them in transit, particularly in environments with weak encryption or monitoring.

In 2020, a breakthrough attack called \textit{SADDNS}~\cite{man2020dns} highlighted DNS’s vulnerability to cache poisoning, despite earlier mitigations like port randomization. SADDNS exploited side-channel vulnerabilities in the \textit{Recursive DNS resolvers} (see Section-~\ref{DNS_section}), allowing attackers to infer randomized port numbers and bypass port randomization protections. This attack demonstrated that DNS was still vulnerable to integrity-based attacks, even with previous defenses in place. A refined version of SADDNS used ICMP responses to further deduce port numbers efficiently~\cite{man2021dns}, indicating that DNS security improvements, while helpful, were insufficient for long-term protection.

These developments underscored the importance of \textit{DNSSEC}, which remains the most robust solution to prevent attacks on the integrity of DNS responses. Yet, the slow adoption of DNSSEC has left DNS exposed, making comprehensive implementation critical to ensuring the security of the protocol. The challenge we will be addressing is that of Islands of Security (see Section-~\ref{DNS_section}). 
% We propose a novel solution built on existing standards to assist in the adaptation of DNSSEC.

In this paper, we propose an extension to DNSSEC, which bridges the gap between islands using existing proven decentralized techniques. We believe these techniques will aid DNSSEC adoption, while being oblivious to the registrar's rate of adoption.

The rest of this paper is arranged as follows: Section~\ref{DNS_section} explains all the terminology used later in the work. Section~\ref{Related} describes the related works on DNS and its extensions. Section~\ref{Proposal}, describes our proposal on bridging the gaps in DNSSEC using TLS. Finally, we conclude our work in Section~\ref{conclusion} with some future directions.

\section{DNS Taxonomy} 
\label{DNS_section}
DNS operates through a hierarchical system typically comprising \textit{Root}, \textit{Top-Level Domain} (TLD), and \textit{Authoritative} nameservers~\cite{rfc1034}. The \textit{Internet Corporation for Assigned Names and Numbers} (ICANN) coordinates the management of the root zone, though the operation of the root nameservers themselves is handled by independent organizations. In this hierarchy, each nameserver is responsible for a portion of the DNS namespace, known as a \textit{zone}. A \textit{zone} refers to a distinct administrative segment, typically corresponding to a domain or subdomain, that is managed and maintained as a single unit. \textit{Registries} are organizations responsible for managing these authoritative databases of domain names within specific TLDs, including the TLD nameservers. \textit{Registrars} act as intermediaries, facilitating domain name purchases and registrations for individuals and organizations, known as \textit{registrants}, who may host their authoritative DNS nameservers either independently or through third-party services~\cite{rfc1591}.

A \textit{Stub Resolver} is the end-user's system-level local DNS client responsible for initiating domain name translation requests. It forwards the request to a \textit{Recursive Resolver}, which is typically hosted locally by the end-user's network administrator/ISP, or hosted publicly by a third-party. The recursive resolver performs iterative resolution, where it queries DNS nameservers in a hierarchical manner, starting from the Root nameserver, moving through the TLD nameserver, and finally reaching the Authoritative nameserver, which holds the requested domain information. Each nameserver provides a referral to the next, narrowing down the search until the authoritative nameserver returns the required DNS record.

Every DNS request is associated with a 16-bit \textit{Transaction ID} (called \texttt{TXID}) and a 16-bit \textit{source port}, both serving crucial roles. The \texttt{TXID} ensures that the client can correctly match DNS queries with their responses, which is especially important when multiple queries are in progress. The source port helps enhance security by randomizing the port number for each query, making it harder for attackers to predict and tamper with responses, while also enabling load balancing across DNS nameservers.

\begin{figure}[htbp]
\centerline{\includegraphics[width=0.5\textwidth]{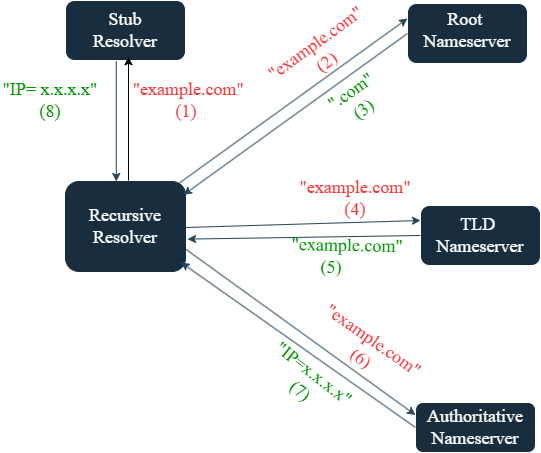}}
\caption{DNS Hierarchy}
\label{fig1}
\end{figure}

Figure~\ref{fig1} illustrates the process of domain name resolution over the DNS nameserver hierarchy. The stub resolver initiates a request for the IP address of \texttt{www.example.com}. The recursive resolver first queries the root nameserver, which refers it to the appropriate TLD nameserver (e.g., \texttt{.com}), and then the TLD nameserver directs it to the Authoritative nameserver for \texttt{example.com}. Once the Authoritative nameserver responds with the IP address for the requested domain (\texttt{example.com}), the recursive resolver verifies its validity, caches the response, and forwards the IP address to the stub resolver, completing the lookup process. The cached response can be used for subsequent queries to resolve the same domain name until the Time To Live (TTL) of the response expires.

To ensure that DNS responses are both authentic and unaltered, DNSSEC implements cryptographic records that work alongside the standard DNS resolution process. These records help validate the integrity of DNS data, guaranteeing that the information received by the resolver hasn’t been tampered with during transit.

DNSSEC introduces three key records to ensure the integrity and authenticity of DNS responses:
\begin{description}
\item [\texttt{DNSKEY} records]
The \texttt{DNSKEY} record contains the public keys used in DNSSEC, and these keys come in two types:

\begin{itemize} \item Zone Signing Key (\texttt{ZSK}) – Used to sign the zone’s resource record sets (\texttt{RRset}), generating the associated \texttt{RRSIG} records.
\item Key Signing Key (\texttt{KSK}) – Used to sign the \texttt{DNSKEY} record itself, providing an additional layer of trust. \end{itemize}

All further references to DNSKEY unless specified will refer, only to the \texttt{ZSK} as it is directly responsible for signing the \texttt{RRset}. The private keys for both \texttt{ZSK} and \texttt{KSK} are held securely by the zone’s administrator, while the corresponding public keys are published in the \texttt{DNSKEY} record. Resolvers use the public \texttt{ZSK} to verify the \texttt{RRSIG} signature of the \texttt{RRset}, ensuring that the records have not been tampered with.

\item [\texttt{RRSIG} records]
The \texttt{RRSIG} record contains the cryptographic signature for an \texttt{RRset}, verifying its authenticity. This signature is generated by the authoritative DNS server using the private \texttt{ZSK} to sign a cryptographic hash of the \texttt{RRset}. When the resolver receives the \texttt{RRSIG}, it retrieves the corresponding public \texttt{ZSK} from the \texttt{DNSKEY} record to verify that the \texttt{RRset} has not been altered.

\item [Delegation Signer (\texttt{DS}) records]
\texttt{DS} records create the trust chain between a parent and child DNS zone. The \texttt{DS} record is a cryptographic hash of the child zone’s \texttt{DNSKEY} using the private \texttt{KSK}, stored in the parent zone. When a resolver queries a child zone, it uses the \texttt{DS} record from the parent zone to authenticate the child zone’s \texttt{DNSKEY}, ensuring that the key hasn’t been tampered with and maintaining the integrity of the chain of trust.

\end{description}
    % \item[DNSKEY records] Public keys used by clients to verify DNSSEC signatures.
    
\begin{figure}[tbp]
\centerline{\includegraphics[width=0.5\textwidth]{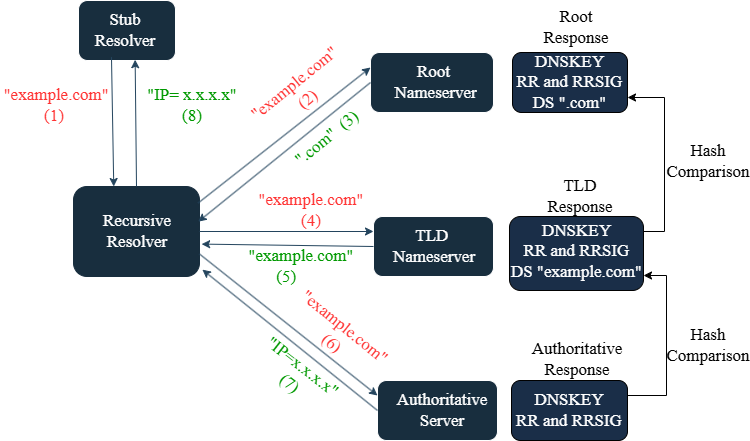}}
\caption{DNSSEC Hierarchy}
\label{fig2}
\end{figure}

DNSSEC builds integrity checks into DNS, with each response now additionally sending respective \texttt{RRSIG}, \texttt{DNSKEY}, and \texttt{DS records} along with the \texttt{RRset}, as shown in Figure~\ref{fig2}. The figure is enumerated to represent the sequence of events, where the \texttt{DNSKEY} is validated against the parent’s \texttt{DS record} (depicted by the hash comparison arrows), ensuring that the resolver is communicating with the correct nameserver and that the \texttt{DNSKEY} has not been tampered with, thus establishing a trust chain. In the case of the root nameserver, where there is no higher level of hierarchy, trust anchors (\texttt{DS records}) present in the resolver are used to validate the root nameserver's \texttt{DNSKEY}. Subsequently, the \texttt{DNSKEY}, along with the \texttt{RRset}, is utilized to calculate a cryptographic hash, which is then validated against the \texttt{RRSIG} provided in the response.

In this paper, fully deployed DNSSEC refers to domains where all levels along the path from root to authoritative nameserver (each parent to child) have properly implemented DNSSEC, including the presence of \texttt{DNSKEY}, \texttt{RRSIG} and \texttt{DS records} at each level. Conversely, partially deployed DNSSEC refers to domains where the \texttt{DS} record is absent in the parent zone (such as the TLD or Second-Level Domain), weakening the chain of trust.

% A critical challenge in receiving integrity guarantees from DNSSEC is that we need a fully deployed DNSSEC. Even if DNSSEC is properly implemented at a level, it offers limited or no protection if the parent zone (or any other zone in the hierarchy) fails to support DNSSEC. This issue, known as the \textbf{Islands of Security} \cite{rfc4033} problem, arises because isolated DNSSEC implementations lack the necessary upstream validation to be fully effective. Furthermore, the increased complexity of managing cryptographic keys, signing records, and validating signatures has contributed to low adoption rates of DNSSEC across domains as studied in \cite{yang2010deploying}.
A critical challenge in achieving full integrity guarantees from DNSSEC is the requirement for fully deployed DNSSEC across the DNS hierarchy. Even if DNSSEC is properly implemented at one level, it offers limited or no protection if the parent zone (or any other zone in the hierarchy) does not support DNSSEC. This issue, referred to as the Islands of Security problem~\cite{rfc4033}, occurs because isolated DNSSEC implementations lack the upstream validation necessary for complete security. Additionally, the increased complexity involved in managing cryptographic keys, signing records, and validating signatures has contributed to the low adoption rates of DNSSEC across domains~\cite{yang2010deploying}.

In contrast to DNSSEC, which primarily ensures the integrity and authenticity of DNS records, \textit{DNS-over-TLS (DoT)}~\cite{rfc7858} enhances DNS security by addressing confidentiality. DoT encrypts DNS queries and responses by routing them through a secure Transport Layer Security \textit{(TLS)} connection, preventing eavesdropping and man-in-the-middle attacks.

In DoT, after a client sends an initial request to a DNS nameserver, a TLS handshake is performed to establish a secure communication channel. During this handshake, the DNS nameserver sends a \textit{TLS certificate}, which is verified by the client to ensure the nameserver's identity. This certificate is typically issued by a trusted Certificate Authority (CA), similar to the process used in HTTPS connections. By verifying the nameserver’s identity through the certificate, the client can confirm that it is communicating with a legitimate DNS nameserver, mitigating risks of impersonation or DNS spoofing.

Once the TLS connection is established and the nameserver's identity is authenticated, DNS queries and responses are exchanged securely, encrypted to maintain confidentiality. While DoT does not provide the cryptographic integrity checks of DNSSEC, it ensures that DNS communications are encrypted and authenticated. 

% DNSCurve (Unclear about inclusion)
% Another 

% 

In summary, while DNSSEC addresses integrity and authenticity-related vulnerabilities in the DNS, its adoption has been hindered by the necessity of full-chain deployment and the operational challenges it introduces. This paper explores these issues in depth and proposes solutions to overcome the barriers to widespread DNSSEC adoption.

\section{Related Work}
\label{Related}
The challenge of Islands of Security in DNSSEC has been a significant barrier to its widespread adoption. Islands of Security arise when atleast one nameserver in the middle of the DNS trust chain does not implement DNSSEC, or when a parent nameserver fails to maintain up-to-date \texttt{Delegation Signer (DS) records} for a child nameserver. This breaks the authentication chain, rendering the DNSSEC deployment ineffective, as a domain’s \texttt{DNSKEY} cannot be validated.

A study on the role of registrars~\cite{chung2017understanding} revealed that while 95\% of generic Top-Level Domain (gTLDs) and approximately 50\% of country-code Top-Level Domain (ccTLDs) are DNSSEC-compliant, many registrars managing these registries either do not support \texttt{DS record} submission or lack proper validation checks for uploaded signature records. Of the top 20 registrars, only 11 support DNSSEC, with just 8 providing web-based \texttt{DS record} submission and 3 relying on insecure email-based DS submissions. Furthermore, only 0.31\% of \texttt{.org} domains are fully DNSSEC compliant, while over 1\% have deployed DNSSEC partially. Though these statistics improved (as suggested in~\cite{roth2019tracking}) some of the registrars are still not validating the \texttt{DS record} supplied by the child. This study also shows a higher willingness to adopt DNSSEC. These figures suggest that while some domains are prepared to implement DNSSEC to secure their DNS responses, the registrars managing these domains have been slow to adopt the necessary mechanisms.

To address such longstanding issues, the study~\cite{kim2006resolving} proposed few mechanisms for DNSSEC validation in the absence of full deployment across the DNS hierarchy. The authors proposed a public \texttt{bulletin board} system, where DNS zones could post their \texttt{DNSKEY} information. This bulletin board would provide the necessary authentication to resolvers even when parent zones failed to implement DNSSEC. While the bulletin board system ensured that key information was available for validation, it did not guarantee the authenticity or correctness of the posted keys. The focus was instead on completeness, meaning that the system prioritized making as many DNSKEY records as possible accessible to resolvers, leaving the responsibility of verifying the trustworthiness of these keys to the resolvers themselves. This lack of built-in key verification posed potential risks, as incorrect or malicious keys could be posted without any central authority to authenticate them.
% To overcome the limitations in validating keys for correctness, the concept evolved into 

\begin{figure}[tbh]
\centerline{\includegraphics[width=0.5\textwidth]{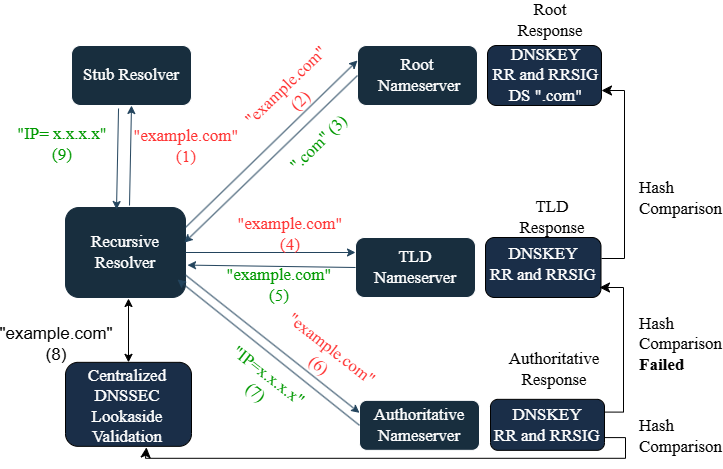}}
\caption{DNSSEC Lookaside Validation (DLV)}
\label{fig3}
\end{figure}

While overcoming this limitation and working towards a practical implementation, this evolved into \textit{DNSSEC Lookaside Validation (DLV)}, introduced in RFC 5074~\cite{rfc5074}. DLV sought to address incomplete DNSSEC deployment by providing an alternative method for validating DNSSEC signatures when the chain of trust was broken due to a missing \texttt{DS record} in the parent zone.
In DLV, a centralized DLV registry acted as a repository for \texttt{DS record}s that would typically reside in the parent zone, as shown in Figure~\ref{fig3}. Instead of relying solely on the DNS hierarchy, DNS resolvers could query this registry for the \texttt{DS record}s of zones, whose parent zones were not signed with DNSSEC or is unable to properly host the \texttt{DS record}s of the child. This provided a temporary workaround for the absence of fully deployed DNSSEC, extending the chain of trust by enabling validators to access \texttt{DS record}s outside the traditional parent-child DNS structure. The implementation of DLV was carried out by the \textit{Internet Systems Consortium} (ISC), hosted in the separate zone \texttt{dlv.isc.org}. This addresses the concerns of the bulletin board approach by having a central entity (ISC) that verifies the legitimacy of the uploaded DNSKEY.

While DLV successfully bridged the gap in DNSSEC deployment, it had significant limitations, particularly its reliance on maintaining a centralized DLV registry and upload verification. As DNSSEC adoption expanded, especially at the root and TLD levels, DLV became less relevant, and the cost of managing the centralized system outweighed its benefits. As a result, ISC decommissioned DLV in 2017. Despite the increased DNSSEC support at the TLD level, registrar adoption remains less than optimal.

\section{Proposed Work}
\label{Proposal}
In this section, we describe our method (in section~\ref{Our work}),  explain the extent of security provided by our work in different situations (in section~\ref{Coverage}), and,  discuss its advantages over traditional DNSSEC implementation (in section~\ref{Discussion}).
\subsection{Building Bridges}
\label{Our work}
In this work, we aim to enhance the security landscape of DNS by reducing the domain owner's reliance on the registrar for DNSSEC compliance. Our approach proposes the use of a TLS-based connection, where a certificate is attached to the IP address of the authoritative nameserver. This certificate is used to verify the legitimacy of DNS responses, thereby introducing an additional layer of security in scenarios where DNSSEC may be incomplete or compromised. For the purpose of this study, we focus on the trust link between the \textit{TLD nameserver} and the \textit{authoritative nameserver}, as this is a frequent point of failure in DNSSEC chains of trust.

\begin{figure}[tbh]
    \begin{subfigure}[tbh]{0.5\textwidth}
        \centering
        \includegraphics[width=\textwidth]{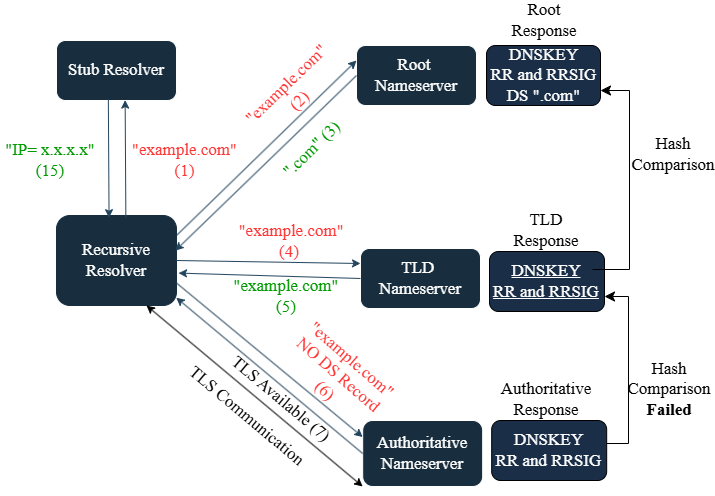}
        \caption{High-level view of Partially deployed DNSSEC and our solution}
        \label{fig4a} 
    \end{subfigure}
    \hfill
    \begin{subfigure}[tbh]{0.5\textwidth}
         \centering
        \vspace{10mm}
        \includegraphics[width=0.7\textwidth]{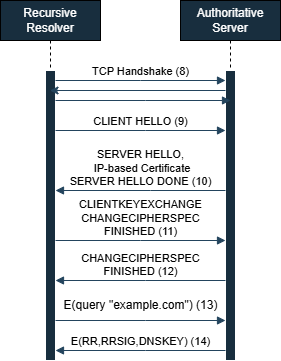}
        \caption{Detailed description of TLS communication between the recursive resolver and the authoritative nameserver in Figure~\ref{fig4a}}
        \label{fig4b}
    \end{subfigure}
    \caption{The Figure~\ref{fig4a} shows the high-level view of DNS hierarchy and presence of keys, while Figure~\ref{fig4b} gives a detailed view into TLS communication when \texttt{DS record} in parent is missing. Numbered labels are used to highlight important steps in the process. The labels in both the diagrams correspond to detailed explanations provided in section~\ref{Our work}. }
    \label{fig4}
\end{figure}

The proposed approach can be summarized in the following steps (as illustrated in Figure~\ref{fig4a},~\ref{fig4b}):

\begin{enumerate}
    \item A DNS request is initiated by the stub resolver and sent to the recursive resolver.
    \item If the request is not available in the recursive resolver's local cache, the resolver forwards the request to the root nameserver.
    \item The root nameserver responds with the \texttt{RR set, RRSIG, DNSKEY}, and the \texttt{DS record} for the child (TLD nameserver).
    \item The recursive resolver verifies the \texttt{DNSKEY} using the \texttt{DS record}, which is part of its trust anchor, and subsequently validates the \texttt{RR} set using the \texttt{RRSIG}. The response typically contains the address of the next link in the hierarchy (i.e., the TLD nameserver), to which the resolver then sends a follow-up request.
    \item The TLD nameserver responds with its corresponding \texttt{RR set, RRSIG}, and \texttt{DNSKEY}. However, in this scenario, the \texttt{DS record} for the authoritative nameserver is \textbf{missing}, which breaks the chain of trust.
    \item Despite the missing \texttt{DS record}, the recursive resolver proceeds by verifying the \texttt{DNSKEY} of the TLD nameserver with the \texttt{DS record} obtained from the root nameserver. It then checks the validity of the TLD's \texttt{RR set}. Aware of the missing \texttt{DS record} for the child (authoritative nameserver), the resolver flags this either by setting a bit in the DNS query or adding information to the additional section.
    \item The authoritative nameserver, upon receiving the query, recognizes that the recursive resolver was not provided with a \texttt{DS record} for its child. If the child domain has not adopted our technique, the authoritative nameserver responds with a regular (unsigned) RR set. However, if the child domain supports our technique, the authoritative nameserver signals the availability of this feature to the resolver.
    \item Upon receiving the indication of DNSSEC support from the authoritative nameserver, the recursive resolver initiates a \textit{TCP handshake} with the authoritative nameserver.
    \item The recursive resolver then begins the \textit{TLS handshake} by sending a \texttt{CLIENT HELLO} message.
    \item The authoritative nameserver responds with a \texttt{SERVER HELLO}, its certificate (which is bound to the \textit{IP address} of the nameserver, rather than the typical certificate bound to a domain name), and the \texttt{SERVER HELLO DONE} message.
    \item The recursive resolver verifies that the IP address of the authoritative nameserver (as received from the TLD nameserver) matches the IP address in the certificate. If the IP addresses do not match, the connection and the DNS request are aborted. If the addresses match, the resolver proceeds by sending the \texttt{ClientKeyExchange}, \texttt{ChangeCipherSpec}, and \texttt{FINISHED} messages.
    \item The authoritative nameserver responds with \texttt{ChangeCipherSpec} and \texttt{FINISHED}, completing the handshake.
    \item The recursive resolver then sends the DNS query to the authoritative nameserver over the encrypted connection.
    \item The authoritative nameserver decrypts the request and responds with an encrypted DNS response, containing the \texttt{RR set}, \texttt{RRSIG}, and \texttt{DNSKEY} for the domain in question.
    \item The recursive resolver decrypts the response and validates the resource records with their signatures and passes it on to the stub resolver.
\end{enumerate}

This method ensures that even when DNSSEC fails to provide full coverage (e.g., missing \texttt{DS record}s), the recursive resolver can still establish a secure and authenticated communication channel with the authoritative nameserver using TLS. By binding the certificate to the IP address of the nameserver, this approach reduces the risks associated with DNS spoofing due to misconfigured or missing \texttt{DS record}s. It also facilitates gradual DNSSEC deployment by minimizing reliance on registrars for \texttt{DS record} management, improving security without requiring immediate, full DNSSEC adoption.

\subsection{Scope}
\label{Coverage}
Though we have discussed bridging the gap between the TLD and Authoritative nameserver, we now aim to generalize the coverage of our work. The two key situations which can occur are as follows: 

\begin{description} \item[Zero-gap] This situation occurs when all DNS entities are DNSSEC-enabled, yet gaps arise due to improper management of the \texttt{DS record}. In this case, while \texttt{DNSKEY} and \texttt{RRSIG} confirm the integrity of the records, the absence of a properly managed \texttt{DS record} disrupts the authentication chain. Our proposal specifically addresses these scenarios by replacing the missing \texttt{DS record} with TLS certificates based on IP verification, thereby ensuring the nameserver's authenticity and restoring the trust chain.

\item[Non-zero gap] This refers to situations where one or more hierarchically consecutive DNS nameservers have not implemented DNSSEC, resulting in islands of security at those levels. In these scenarios, our solution provides protection similar to  DoT, guaranteeing the integrity of data during transit and authenticating the nameserver being contacted. However, since DNSSEC is not applied, the integrity of the DNS records themselves cannot be verified. \end{description}

\subsection{Discussion}
\label{Discussion}
Our approach presents a \textit{decentralized} solution to address the lack of DNSSEC support in parent zones. Similar to the DLV, we propose a mechanism that simplifies DNSSEC adoption by removing the dependency on the parent nameserver's compliance, which is often beyond the domain owner's control. By allowing DNSSEC validation to occur independently of parent nameserver support, our solution facilitates broader adoption of DNSSEC without requiring immediate or full compliance across the DNS hierarchy. However, unlike the DLV and bulletin board approaches, which store records in a central location and place trust in the entity responsible for managing them, our work relies solely on the decisions of the nameserver. The burden of trust is not centralized, and the responsibility for implementation rests only on the nameserver that implements this feature.

\textit{Advantages}: Our approach leverages well-established protocols, particularly TLS and DNSSEC, thereby inheriting their robustness, security features, and widespread adaptability. This allows us to focus on enhancing DNS security without the need to address potential vulnerabilities or implementation challenges within the TLS protocol itself.

\textit{Overhead}: While our method introduces additional computational overhead and may slightly increase latency due to extra round-trip times (RTTs) and cryptographic operations, these trade-offs are minimal compared to the significant security benefits provided. The enhanced protection against a wide range of DNS attacks offered by our approach outweighs the associated cost of three additional RTTs and cryptographic operations.

% \textbf{Novelty:}
\textit{Contrast with DoT}: Although our approach shares similarities with DNS over TLS (DoT), it offers superior performance and security. Unlike DoT, which requires establishing a TLS connection at every level of the DNS hierarchy, our method reduces the need for repeated TLS handshakes, thereby minimizing latency. Furthermore, DoT primarily focuses on encrypting DNS queries and responses to ensure confidentiality, but it does not provide comprehensive integrity guarantees. 
% \comments{red}{Check the following }
Our solution, by contrast, aims to deliver DNSSEC-level cryptographic integrity assurance even in scenarios where \texttt{DS record}s are not properly managed, ensuring protection against DNS tampering in fully DNSSEC-enabled environments. In cases where one or more levels in the DNS hierarchy are not DNSSEC-enabled, our solution provides security guarantees that bridge the gap between the comprehensive protection of DNSSEC and the encryption-focused security offered by DoT, thereby maintaining higher integrity across both partially secure and fully secure setups. 

% However, the ease of obtaining a certificate bound to IP Address is still to be explored. 

% Our work is also better than DoT, which similarly uses TLS as in DoT every connection to every nameserver required 4 RTTs while ours has to make only 1 4 RTT step for where the gap is to be bridged.

\section{Conclusion And Future Scope}
\label{conclusion}
In this paper, we addressed the challenges posed by partially deployed DNSSEC and the resulting Islands of Security. We proposed a decentralized solution that leverages TLS and IP-based TLS certificates to authenticate the child nameserver using the IP address provided by the parent nameserver, effectively bridging the authenticity gaps in these Islands of Security. This approach eliminates the dependency on the parent nameserver for correctly hosting and validating the child's \texttt{DS record}. Our solution provides immediate security benefits for DNSSEC-compliant domains while encouraging broader adoption of DNSSEC. Additionally, it offers registrars more flexibility, allowing them time to implement efficient and reliable mechanisms for \texttt{DS record} acquisition and management. Although our method introduces some computational overhead and increased latency due to cryptographic operations and additional round-trip times (RTTs), the enhanced security it offers significantly outweighs these trade-offs.

 Implementation of this technique would require modifications to the resolver and nameserver software such as BIND~\cite{bind} and Unbound~\cite{unbound}. Experimentation on the implementation will help us understand of practical difficulties, if any. Also the ease of obtaining the IP-based digital certificates and costs of enabling this technique on the organizations are to be studied.

\balance
\bibliographystyle{IEEEtran}
\bibliography{references}
\end{document}